\documentclass[twocolumn]{revtex4}
\usepackage{dcolumn}
\usepackage{graphicx}
\usepackage{graphicx,amssymb,amsmath}
\usepackage{natbib}
\usepackage{bm}
\begin{document}

%%%%%%%%%%%%%%%%%%%%%%%%%%%%%%%%%%%%%%%%%%%%%%%%%%%%%%%%%%%%%%%
%     This is version seven.
%     Last changed on June 2, 2003 by AJL.
%%%%%%%%%%%%%%%%%%%%%%%%%%%%%%%%%%%%%%%%%%%%%%%%%%%%%%%%%%%%%%%

\title{Dynamics of rigid and flexible extended bodies in viscous films and membranes}
\author{Alex J.~Levine$^{1,2}$, T.B.~Liverpool$^{1,3}$ and
  F.C.~MacKinstosh$^{1,4}$}
\affiliation{$^{1}$The Kavli Institute for Theoretical
  Physics, University of California, Santa Barbara CA 93106}
\affiliation{$^{2}$Department of Physics, University of Massachusetts,
  Amherst Amherst MA 01060} 
\affiliation{$^{3}$Department of Applied
  Mathematics, University of Leeds, Leeds, LS2 9JT, United Kingdom }
\affiliation{$^{4}$Division of Physics \& Astronomy, Vrije Universiteit 1081
  HV Amsterdam, The Netherlands} %\email[]{fcm@nat.vu.nl}

\date{\today}
\begin{abstract}
%%FCM:
We study the dynamics of extended rod-like bodies in (or associated
with) membranes and films. We demonstrate a striking difference
between the mobilities in films and bulk fluids, even when the
dissipation is dominated by the fluid stress: for large inclusions we find that
rotation and motion perpendicular to the rod axis exhibits purely
local drag, in which the drag coefficient is algebraic in the
rod dimensions.  We also study the dynamics of the {\em internal}
modes of a semiflexible inclusion and find two dynamical regimes in
the relaxation spectrum.

\end{abstract}

\pacs{PACS numbers: 83.10.-y, 87.68.+z, 87.17.Dg}

\maketitle

The mobility of inclusions in membranes is a fundamental physical
parameter controlling a number of cellular processes. Since
inclusions such as proteins~\cite{Stein,Spooner} or
``rafts''~\cite{Simons,Pralle} can in many cases be large compared
with the constituent lipids of the membrane, they can be viewed as
macroscopic objects moving in a continuum fluid environment. The
dynamics of these objects in thin films has already been shown to
be surprisingly subtle, leading to confusion in the early
literature on protein diffusion in cell membranes~\cite{Hughes}.
This controversy was clarified by Saffman~\cite{Saffman}, who
noted that motion of objects in or attached to a (two dimensional)
membrane is never strictly two-dimensional since in-plane momentum
induces flows in the surrounding bulk (three-dimensional) fluids
due to a viscous coupling of the interface/membrane to the
surrounding fluids. That coupling introduces a new length-scale
$\ell_0=\eta_{\rm m}/\eta_{\rm f}$ determined by the ratio of
membrane and fluid viscosities, $\eta_{\rm m, f}$. This length
determines the degree to which the dissipation is predominantly
two- or three-dimensional
\cite{Saffman,Hughes,Lubensky,Ajdari,Helfer:2000,Levine:02}. As a
result, the drag coefficient on a small object in a membrane is a
nonlinear function of both its size and the membrane viscosity.
For example, the diffusion coefficients of small proteins (linear size
$< \ell_0$) in membranes have a weak (logarithmic) dependence on
their size.

Here, we examine the motion of rod-like inclusions in viscous
films and membranes, as a representative example of a general
scheme for the calculation of 
%%FCM:
mobilities of arbitrary extended bodies, which we also present. The
generalization of this problem to motion in viscoelastic films is
straight-forward. Our work is motivated in part by prospect of
computing mobilities of proteins and lipid rafts in lipid bilayers
\cite{Stein,Spooner}, as well as by recent experiments that have
demonstrated the possibility of making quantitative rheological
measurements of viscoelastic films resembling cellular structures
such as the {\em actin cortex} \cite{Helfer:2000} using tracer
particle fluctuations (membrane microrheology). 
%%ADDED BY AJL 
Given its flexibility, this formalism can be adapted to the computation of the mobility of 
highly complex and irregular interfacial objects such as fractal aggregates\cite{Warren:94}. 
Finally, 
%% END OF ADDITION.
the driven motion of
rods in viscous/viscoelastic films has also been used to determine
rheological properties, \emph{e.g.}, of monolayers \cite{Z:00},for
which the approach we develop here is important. 
%%FCM:
We also analyze the undulatory motion of semiflexible
polymers at a membrane surface.

Our calculation of the hydrodynamic drag on a {\em rigid} 
rod of length $L$, gives two principal results: (i) for small objects
(specifically, $L\ll\ell_0$), the drag coefficients become independent
of both the rod orientation and aspect ratio; and (ii) for larger rods
of high aspect ratio, $\zeta_\perp$ becomes purely linear in the rod
length $L$---\emph{i.e.}, the drag becomes purely local. In contrast,
we find that the well-established three-dimensional result
$\zeta_\parallel=2\pi\eta/\ln(AL/a)$ applies for parallel motion in
the film, provided that $L\gg\ell_0$. Here, however, the effective rod
radius becomes $\ell_0$ rather than the physical radius $a$, when
$a\ll\ell_0$.  Closely related to (ii), we find that the rotational
drag (equivalently diffusion constant) depends purely algebraically on
the rod length. 

We also calculate the relaxation spectrum of undulation modes of wavelength
$q^{-1}$ of a {\em semiflexible} inclusion of rigidity $\kappa$ and
find a dynamic relaxation time $\tau(q)$ which crosses over from $\tau
\sim q^{-3}$ on modes with wavelength less than $\ell_0$ to $\tau \sim
q^{-4}$ for those longer than $\ell_0$. 
%%FCM:
The latter result represents purely local drag, with 
no additional logarithmic dependence in
$\tau(q)$, as there is for bending fluctuations in a bulk fluid.

The starting point of our analysis will be the response of the two
dimensional membrane fluid to an in-plane force distribution. This
is calculated by solving the coupled equations for in-plane and
out of plane fluid motions taking into account incompressibility
\cite{Levine:02} of both the bulk and the membrane leading to an
expression for the in-plane velocity $v_\alpha \left( {\bf
x}\right)$ at position ${\bf x}$ resulting from a point force 
$f_\beta \left({\bf x'} \right)$ 
at {\em another} point ${\bf x'}$:
$ v_\alpha \left( {\bf x}\right) =
\alpha_{\alpha \beta}\left({\bf x} - {\bf x}' \right) f_\beta
\left({\bf x}' \right)\,
$.
The response function $\alpha_{\alpha \beta}\left({\bf x}
\right)$ is given in closed form as
\begin{equation}
\label{response-tensor}
 \alpha_{\alpha \beta}\left({\bf x}
\right)= \alpha_\parallel \left(\left|{\bf x} \right| \right)
\hat{x}_\alpha \hat{x}_\beta +\alpha_\perp \left(\left| {\bf
x} \right| \right) \left[ \delta_{\alpha \beta} - \hat{x}_\alpha
\hat{x}_\beta \right],
\end{equation}
where the scalar functions $\alpha_\parallel, \alpha_\perp$ of the
distance between the point of the force application and the
measurement of the velocity field are given by:
\begin{equation}
-4\eta_m i\omega\alpha_{\|}(x,\omega) ={{\bf H}_1(z)\over
z} -{2\over \pi z ^2}-{Y_0(z)+Y_2(z)\over 2} \label{para-def}
\end{equation}

\begin{eqnarray}
\nonumber && -4\eta_m i\omega \alpha_\perp (x,\omega)=\\
&& {z{\bf H}_0(z)-{\bf
H}_1(z)\over z} +{2\over \pi z^2}-{Y_0(z)-Y_2(z)\over 2}, \label{perp-def}
\end{eqnarray}
where the ${\bf H}_\nu$ are Struve functions\cite{Abramowitz:64},
and the $Y_\nu$ are Bessel functions of the second kind. Here,
$z= \left| {\bf x} \right|/ \ell_0$ is the distance between the
point of force application and the membrane velocity response
measured in the flat film in units of $\ell_0$.

To parameterize this geometry of the rod of length $L$ and
circular cross-section $a$, we define the dimensionless aspect
ratio $\rho= L/a$. There are three independent drag coefficients
to determine in the problem. The in-plane translational mobility
tensor, $\mu_{\alpha \beta}$ (and consequently, its inverse, the
drag $\zeta_{\alpha \beta} = \mu^{-1}_{\alpha \beta}$) is defined
by
\begin{equation}
\label{alpha-def} v^{\mbox{rod}}_{\alpha} = \mu_{\alpha \beta}
F_\beta^{\mbox{rod}},
\end{equation}
where $v^{\mbox{rod}}_{\alpha}$ is the $\alpha^{\rm th}$ component of
the velocity of the rod and $F_\beta^{\mbox{rod}}$ is the $\beta^{\rm th}$
component of the total force applied to the rod (at its center)
and $\alpha,\beta \in \{1,2\}$.  In-plane rotational symmetry and
inversion of the rod imply that the mobility tensor has the form:
\begin{math}
\label{alpha-two-parts} \mu_{ij}= \mu_{||} \hat{n}_i \hat{n}_j +
\mu_{\perp} \left( \delta_{ij} - \hat{n}_i \hat{n}_j \right).
\end{math}
Here $\mu_{\perp}$ and $\mu_{\|}$ are the mobility of the rod
dragged perpendicular to and parallel to its long axis 
%% AJL ADDED
(oriented along the $\hat{n}$ direction) 
%%  END OF ADDITION
respectively.
In addition to these two independent translational mobilities, there
is also one rotational mobility, $\mu_{\rm rot}$ linking the angular
velocity of the rod to the torque applied to that rod about its center
of inversion symmetry.

We solve the problem by two complementary methods useful for the
regimes of small and large aspect ratios respectively.  For small
$\rho$, we use a 2-d analogue of the Kirkwood
approximation~\cite{Doi} to model the continuous rod by a series
of discs subject to point forces at their centers. This method
becomes rather cumbersome when $\rho \gg 1$, but here we may
proceed by a second approximation that assumes the rod to be
infinitely thin. In both cases, we restrict our attention to the
limit $a \ll \ell_0$.  Below, we illustrate both methods by a
calculation of the transverse drag coefficient with the
understanding that the longitudinal and rotational drag proceed
analogously.

The linearity of the underlying low-Reynolds number hydrodynamics
allows to use superposition.  Specifically, we replace the rod of
length $L$ and cross--sectional radius $a$ by a set of $N+1$ disks
of radius $a$ and interdisk separation $b$ chosen so that the
total length of the rod is preserved, {\i.e.\/} $L = Nb + 2 a $.
We choose the number of disks to be maximal consistent with a
given aspect ratio and the non-interpenetrability of the disks.

Our strategy for computing the drag on the rod involves setting
the rod in uniform motion with unit velocity by imposing some set
of forces ${\bf f}^{(i)}$, $i = 1,\ldots,N+1$ on the $N+1$ disks
making up the rod.  We can use the response function to compute
the velocity field for a given collection of point forces
\begin{math} \label{drag} v^{(i)}_{\parallel,\perp} =
\sum_{i=1}^{n+1}
\alpha_{\parallel,\perp}^{(ij)}f^{(j)}_{\parallel,\perp}
\end{math}. However, we must demand that all the disks have the
{\it same\/} velocity and thereby determine the forces applied to
them.  To enforce this constraint, we invert the matrix
$\alpha_{\parallel,\perp}^{(i,j)}$. The drag coefficient is then,
\,
\begin{math}
\zeta_{\perp,\|} = \sum_{i,j=1}^{n+1} \left(\alpha^{-1}_{\perp, \|}\right)^{(i,j)}\, .
\end{math}

For large $N$, {\it i.e.\/} high aspect ratio rods, the matrix
inversion becomes difficult. To study that limit, one can assume
an infinitely thin rod. Then, the velocity field at the point
${\bf x}$ due to a continuous distribution of force densities
along the rod, ${\bf f}(x \hat{x})$ that lie along the
$\hat{x}$-axis from $x = -L/2$ to $x = L/2$ takes the form
\begin{equation}
\label{thin-setup} v_\alpha({\bf x}) = \int_{-L/2}^{L/2} \alpha_{\alpha \beta}
\left( {\bf x} - p \hat{x} \right) f_\beta \left( p \hat{x} \right)
dp\, .
\end{equation}
As before, we impose a unit velocity field on the rod and
determine the force density required to effect this result, ${\bf
f} \left( {\bf x} \right) $. The inversion of Eq.~\ref{thin-setup}
proceeds by first expanding the linear force density in Legendre
polynomials $P_n(x)$ writing
\begin{equation}
\label{force-expansion} f(x) = \sum_{n=0}^N c_{2n} P_{2n}(2x/L)
\end{equation}
where the coefficients are as yet unknown. Since the Legendre
polynomials form a complete set on the interval $-1$ to $+1$, any
physical force density can be expressed as in
Eq.~\ref{force-expansion} provided $N$ be taken to infinity. In
practice, we find excellent numerical results even when truncating
this sum to just the first five terms, while taking into account
the symmetry of the force distribution about the center of the
rod.

We now impose the unit velocity condition at a finite set of
points $0\le p_i\le L/2$ along the rod since if we truncate the
Legendre function expansion of the force density at $N$, we can
impose the velocity condition at a maximum of $N$ points without
creating an over-determined system of equations. Thus we require
that
\begin{math}\label{requirement} {\bf v} \left( p_i \hat{x} \right)
=1, {\mbox{for}} \, i = 1,\ldots,N
\end{math}. Now finding the force distribution along the rod
requires only the inversion of an $N \times N$ matrix, ${\mathcal
N}_{ij}$ whose components are defined by
\begin{equation}
\label{matfinal} {\mathcal N}_{ij} =\int_{-L/2}^{L/2} \alpha
\left( \hat{x} p_i - z \hat{x} \right) P_j(2z/L) \, dz.
\end{equation}
Finally, the total force on the rod is found by reconstructing the
force density from its Legendre polynomial expansion and
integrating the resulting expression over the entire rod.  That
force density is given by Eq.~\ref{force-expansion} where the
coefficients, $c_k$ are determined using the rod's unit velocity
condition enforced at a discrete set of points along with the
inverse of the matrix defined in Eq.~\ref{matfinal}. Thus we find
\begin{math}
\label{force-solution} c_k = \sum_{i=1}^N {\mathcal N}^{-1}_{ki}.
\end{math}
Due to the orthonormality of the Legendre polynomials, the total
force is given entirely by the coefficient of the zeroth Legendre
polynomial,and the total torque in the case of rotations is given
by the coefficient of the first odd Legendre polynomial, $c_1$.
Numerically, we find a variation of less than a percent in the
drag coefficient in all cases, so long as $N>2$. Here, we report
our results for $N=5$.  To check for consistency, we compared our
results from this method and the Kirkwood approximation; for thin
rods, we find excellent agreement.

\begin{figure}[htpb]
\centering
\includegraphics[width=6.0cm]{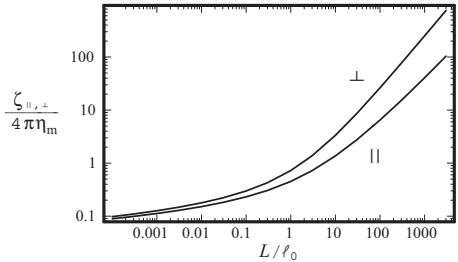}
\caption{Parallel (\protect{$\|$}) and perpendicular $\perp$ drag
coefficients for thin rods ({\em i.e.\/} high aspect ratio) of
various lengths. For short rods both the parallel and
perpendicular drag coefficients approach the Saffman-Delbr\"{u}ck
result.  The divergence between these two drag coefficients at
long lengths reflects the absence of the hydrodynamic
cooperativity in perpendicular case.} \label{resultone}
\end{figure}

The analogous calculation can be made for the drag on the rod
moving in a direction perpendicular to its long axis.  The two
drag coefficients (parallel and perpendicular) converge rapidly
for rods of length $L < \ell_0$, while their difference
grows monotonically
with increasing rod length for $L > \ell_0$. This results stands
in contrast to the case of motion in bulk fluids, where the two
drag coefficients differ by a constant multiple.

We plot the parallel and perpendicular drag coefficients for thin
rods as a function of their length, $L/\ell_0$ in reduced units in
Figure \ref{resultone}. The equality of the two drag coefficients
when all dimensions of the rod are small compared to $\ell_0$
reflects the effective shape and size independence of the
Saffman-Delbr\"{u}ck result. For rods longer than this crossover
length, these two drag coefficient diverge from each other while
they both become more strongly length dependent.

To understand this peculiarity of interfacial drag, we note two
points. First, for long rods ($L > \ell_0$), the parallel drag in
the film is essentially unchanged from the bulk, three-dimensional
drag, in that
\begin{math}
\zeta_{\|} = \frac{2 \pi \eta L}{\ln \left( 0.43 L/\ell_0
\right)},
\end{math}
where the prefactor in the logarithm has been determined to within
$1\%$.  Comparing with the result for drag of a rod in a bulk
fluid, we see that the effective radius of the rod is now of order
$\ell_0$ (for $a\ll\ell_0$). Given that $\ell_0$ corresponds to a
length scale over which interfacial momentum densities flow into
the bulk fluid due to the viscous coupling between the two, the
system does not effectively
resolve length scales smaller than
$\ell_0$ so the small dimension of the rod is replaced by this
length. At large length scales (compared to $\ell_0$) the fluid
velocity field around the rod in parallel drag is the same as for
rod motion in the bulk fluid, {\it i.e.\/} in both cases there
should have been no flow from the plane of the interface into the
subphase.

The case of perpendicular motion of long rods, on the other hand,
is qualitatively very different. Here we find
\begin{math}
\zeta_{\perp} = 2 \pi \eta L.
\end{math}
For this case, the fluid flow field for a rod in bulk is
inconsistent with the flow restrictions imposed by the presence of
the interface. In 3d, there would be a non-vanishing
two-dimensional divergence of velocity field restricted to the
plane of motion of the rod. Now, the in-plane incompressibility of
the interface requires that the fluid velocity field extend over
distances comparable to the rod's largest dimension $L$. The
standard hydrodynamic coupling of portions of the rod, which gives
rise to the logarithm in the drag is not present resulting in a
drag coefficient that is purely linear in rod length. In other
words, the drag is effectively \emph{local} or ``free-draining''
in character. In three dimensions, in contrast, there is a
length-independent ratio of two between the parallel and
perpendicular drag coefficients because neither is free-draining:
$\zeta^{\mbox{3d}}_{\|} = \frac{2 \pi \eta L}{\ln \left( \frac{A
L}{a} \right)}$ and $\zeta_\perp^{\mbox{3d}} =
2\zeta_{\|}^{\mbox{3d}}$.

\begin{figure}[htpb]
\centering
\includegraphics[width=8.0cm]{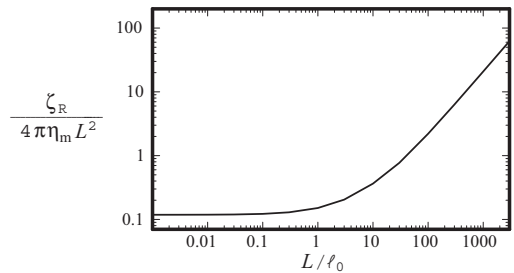}
\caption{The rotational drag coefficient of a rod of infinite
aspect ratio plotted versus the length of the rod. }
\label{rotation}
\end{figure}

\begin{figure}[htpb]
\centering
\includegraphics[width=5.5cm]{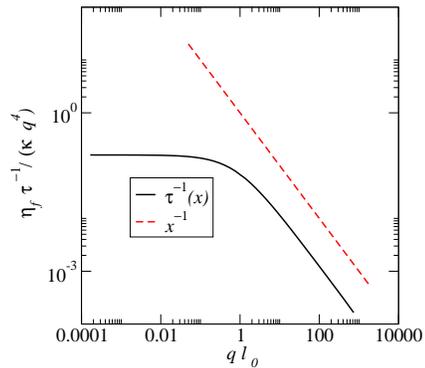}
\caption{The decay rate vs wavevector of transverse fluctuations
of a semiflexible rod embedded in the interface. The decay rate
scaled by $q^4$ is wavevector--independent for purely local drag
(small $q$ regime). Nonlocal hydrodynamic interactions along the
rod reduce the decay rate in a manner analogous to \cite{Brochard}
in the larger $q$ regime.}
\label{decayrate}
\end{figure}

The calculation of the rotational drag coefficient proceeds
analogously to those of the perpendicular and parallel drag
coefficients.  We plot the rotational drag coefficient divided by
$L^2$ for a rod of infinite aspect ratio as a function of the
reduced length in figure \ref{rotation}. The essential feature of
this plot is that rotational drag coefficient scales as $L^2$ for
rods smaller than $\ell_0$ and then as $L^3$ for rods longer than
this natural length.  Thus, we find purely algebraic behavior in
both limits.

Finally, to study the internal mode dynamics of a semiflexible
inclusion, we consider the small transverse fluctuations of a an
infinitely thin, almost rod-like filament oriented along the $x$
axis parameterized by $(x,r(x,t))$. With no slip boundary
conditions on the inclusion we have a dynamical equation for
$r(x)$ to linear order,
\begin{equation}
\partial_t r(x,t) = \int dx' {\bm \alpha}(x-x',0) \cdot 
\left(0, - \kappa \partial_x^4 r \right) + \eta(x,t)
\label{semiflex}
\end{equation}
where the thermal fluctuations $\eta(x,t)$ are chosen to satisfy
the Fluctuation-Dissipation Theorem.  Expanding in modes of
wavenumber $q$, we obtain a relaxation spectrum of transverse
correlations given by
\begin{equation}
\left\langle  \tilde{r}(q,t)  \tilde{r}(q,0)\right \rangle = 
{2 k_B T \over \kappa q^4} \exp \left[ - {t /\tau(q)} \right]
\label{trans_corr}\,.
\end{equation}
where $\displaystyle \tau^{-1}(q) = {2 \kappa q^4 \over
\eta_m}\int_q^\infty {dx \over 2 \pi}{q^2 \over x^2 (x^2-q^2)^{1/2}(
x + \ell_0^{-1})} $. We have obtained a {\em complicated} but {\em
closed form} expression for $\tau(q)$ that is plotted in figure \ref{decayrate}. 
It is simple in the two
limits of $q\ell_0 \ll 1$, $q\ell_0 \gg 1$ where $ \displaystyle
\tau^{-1}(q \ll \ell_0^{-1}) \simeq {\kappa q^4 \over \pi \eta} $
and $\displaystyle \tau^{-1}(q \gg \ell_0^{-1}) \simeq {\kappa q^3
\over 4
\eta_m} $ respectively.  At short scales we find the dimensionally reduced
analogue of the hydrodynamic relaxation spectrum of fluid
membranes ( $D=2$ dimensional manifold embedded in $D+1$
dimensions)~\cite{Brochard} (we consider $D=1$). Here though, the
'long-range' 
hydrodynamic coupling crosses over to a purely {\em
local} friction on the longest length scales \cite{LubenskyNote}.

In summary, we have calculated the translational and rotational
hydrodynamic drag on a rod moving at low Reynolds number in a
viscous film coupled to viscous sub- (and/or) superphase. When the
dimensions of the rod are small ($\ll\ell_0$), the dissipation is
governed primarily by the film, and the drag is insensitive to
orientation and aspect ratio, and only weakly (logarithmically)
dependence on size\cite{Saffman}. We find, surprisingly, that in
the limit of a long rod, the drag for motion perpendicular to the
rod axis exhibits purely local drag per unit length. Thus, in
contrast with motion in 3d fluids, there are no long-range
hydrodynamic effects for long rods, even though the dissipation
occurs entirely in the fluid. A similar observation applies to
rotational motion, and for the relaxation spectrum of the long
wavelength modes of a fluctuating filament or semi-flexible
polymer in a viscous film. This is because, although the
dissipation is governed primarily by the fluid, the film (along
with its assumed incompressibility and no-slip conditions) imposes
a very different boundary condition on the flow from what we would
have in a bulk fluid alone. This added condition not only {\it
increases\/} the drag for perpendicular motion relative to that
without the film present, but results in purely local, or
free-draining drag dynamics. In contrast, since the new boundary
conditions {\em are} consistent with the standard flow field for
parallel motion, we find quantitative agreement with the parallel
mobility in a bulk fluid when the film's viscosity becomes
irrelevant (small $\ell_0$). Finally, we note that the methods
developed here constitute a highly adaptable framework to compute
the mobility of arbitrary extended, irregularly shaped objects
embedded in a viscous membrane or interface.


\begin{thebibliography}{Z:00}
\bibitem{LubenskyNote}That the drag coefficient per unit length
of an oscillating line becomes independent of wavelength in the
long wavelength limit was noted in Ref.\ \cite{Lubensky}.
\end{thebibliography}
\end{document}